\documentclass[useAMS,usenatbib]{mn2e}

\def\bea{\begin{eqnarray}}
\def\ena{\end{eqnarray}}

\usepackage[dvips]{graphicx}
\usepackage{amssymb}

\title{Observing gravitational wave bursts  in pulsar timing measurements}
\author[M. S. Pshirkov,  D. Baskaran and K.A. Postnov]{M. S. Pshirkov$^{1}$\thanks{E-mail:
pshirkov@prao.ru},  D. Baskaran
$^{2,3}$\footnotemark[1]\thanks{E-mail:
Deepak.Baskaran@astro.cf.ac.uk } and K.A.
Postnov$^{4}$\footnotemark[1]\thanks{E-mail:
kpostnov@gmail.com}\\
$^{1}$Pushchino Radio Astronomy Observatory, 142290 Pushchino, Russia\\
$^{2}$School of Physics and Astronomy, Cardiff University, Cardiff CF24 3AA, UK\\
$^{3}$Wales Institute of Mathematical and Computational Sciences, Swansea SA2 8PP, UK\\
$^{4}$Sternberg Astronomical Institute, 119992 Moscow, Russia}

\onecolumn
\begin{document}

\date{}


\maketitle

\label{firstpage}

\begin{abstract}
We propose a novel method for observing the gravitational wave
signature of super-massive black hole (SMBH) mergers. This method
is based on detection of a specific type of gravitational waves,
namely gravitational wave burst with memory (BWM), using
pulsar timing. We study the unique signature produced by BWM in
anomalous pulsar timing residuals. We show that the present day
pulsar timing precision allows one  to detect BWM due to SMBH
mergers from distances up to $1~\rm{Gpc}$ (for case of equal mass
$10^8~M_{\odot}$ SMBH). Improvements in precision of pulsar timing
together with the increase in number of observed pulsars should
eventually lead to detection of a BWM signal due to SMBH merger,
thereby making the proposed technique complementary to the
capabilities of the planned LISA mission.
\end{abstract}




\section{Introduction}
\label{SectionI}

The prospects of detecting gravitational waves (GWs) in the coming
decade are  looking ever more promising
\citep{Grishchuk2001,Cutler2002,Schutz2009}. There is currently a
considerable experimental effort to detect gravitational waves in
a wide range of frequencies. At high frequencies
$\nu\sim10^{1}-10^{3}~{\rm Hz}$ the ground based laser
interferometric detectors, such as the currently operational LIGO
\citep{LIGO} and VIRGO \citep{VIRGO}, especially in their ``Advanced"
configuration, should be able to observe GWs
from a wide range of astrophysical sources. The planned space
interferometer LISA \citep{LISA} would be able to observe GWs in
the region of frequencies $\nu\sim10^{-4}-10^{-1}~{\rm Hz}$. At
intermediate range of frequencies $\nu\sim10^{-8}-10^{-6}$ pulsar
timing measurements are a very strong tool for observing the GW
signature. Several pulsar timing array (PTA) projects, such as
PPTA \citep{PPTA}, EPTA \citep{EPTA} and NANOGrav \citep{NANOGrav}
are collecting data and yielding an unprecedented and every
increasing sensitivity in intermediate frequency range.
Furthermore, the implementation of the planned Square Kilometer
Array (SKA) \citep{SKA} radio telescope would further improve the
sensitivity of pulsar timing measurements to gravitational waves.
Finally, at the lowest  frequencies $\nu\sim10^{-18}-10^{-16}~{\rm
Hz}$ considerable efforts are being made  to measure the imprints
of relic gravitational waves in the temperature and polarization
anisotropies of the cosmic microwave background (CMB) \citep{keating2006}.

In what follows we shall consider pulsar timing, in
particular timing of millisecond pulsar (MSPs), as a tool to study
gravitational waves (see \citep{Hobbs2005,Jenet2005} for a recent
overview). A GW passing between the pulsar and the observer on
Earth will leave a modulating signature on the observed period of
the pulsar. For this reason, the analysis of pulsar timing
residuals gives a unique opportunity to observe GWs
\citep{Estabrook1975,Sazhin1978,Detweiler1979,Bertotti1983,Kopeikin1997}.
A particularly attractive method for detection is to
cross-correlate the timing signal from different pulsars
\citep{Hellings1983}. This method led to the proposal for a PTA
\citep{Romani1989,Foster1990}, which is aimed at observing  a
number of pulsars distributed on the sky  at regular time
intervals over a long time span (see \cite{Verbiest2009} for
recent review).

Current pulsar timing observations are yet to detect gravitational
waves. However, current upper bounds on the stochastic background
of GWs $\Omega_{gw}h^2 <2\cdot10^{-8}$ \citep{Jenet2006} (in terms
of the ratio of energy density per unit logarithmic frequency
interval to the cosmological critical density at frequency
$\nu=1/8{\rm yrs}$) already place interesting limits on cosmic
string models. Moreover, in the near future, it is likely that the
GW background of coalescing extragalactic supermassive black hole
binaries and possibly the background of relic gravitational waves
will be discovered \citep{PPTA}. Apart from the stochastic
backgrounds, pulsar timing also has the potential to observe GWs
from individual sources. In particular, there has been
considerable interest in GWs emitted by massive black hole
binaries systems \citep{Jenet2004,Sesana2009}.

In the present work we shall be interested with individual sources
 which emit gravitational waves of a particular class, namely
gravitational wave bursts with memory (BWM)
\citep{Zel'dovich1974,Smarr1977,Kovacs1978,Bontz1979,Braginsky1985,Braginsky1987},
and analyze the prospects for their detection in pulsar timing
measurements. In general, BWM are characterized by a rise in the
gravitational wave field $h_{ij}^{TT}$ from zero, followed by
oscillatory behaviour for a few cycles and settling down at a
final non-zero value $\Delta h_{ij}^{TT}$ after a characteristic
time $\delta t$ known as the duration of the burst
\citep{Braginsky1987}. The permanent change $\Delta h_{ij}^{TT}$
is usually referred to as the burst's ``memory". BWMs are produced
in situations where there is a net change in the time derivatives
of multipole moments characterizing the system, for example during
flyby of massive bodies on hyperbolic trajectories or an
asymmetric supernova explosion. A BWM where the net change of the
moments is attributed to the gravitons released during the burst
is known as the Christodoulou effect
\citep{Payne1983,Christodoulou1991,Thorne1992,Blanchet1992}.
Whatever the physical reason, the existence of the permanent
offset $\Delta h_{ij}^{TT}$ in BWMs makes them particularly
interesting for pulsar timing studies. As we shall show below, the
burst's memory leads to a characteristic signature in pulsar
timing residuals which accumulates linearly with time.

The structure of the paper is as follows. We begin in
Section~\ref{SectionII} with a simple order-of-magnitude
estimation of the potential sources of BWM, and compare them with
the expected sensitivity of pulsar timing measurements. In this
section, we study the typical amplitudes and possible event rates
for potential signals. The simplistic analysis shows that, in the
context of pulsar timing, the most observationally promising
sources of BWM are extragalactic coalescing supermassive black
holes binaries. In Section~\ref{SectionIII}, we study the
observational prospects for the  detection of the BWM signal in
significant detail. Using a simple analytical model for a BWM
signal we calculate the pulsar timing residual. Following this, we
evaluate the expected signal-to-noise ratio, and the expected
number of BWM events that could be observed in a typical pulsar
timing experiment. Finally, we conclude in Section~\ref{SectionV}
with an overview and discussion of the main results of this work.


\section{Potential sources of gravitational wave bursts with memory}
\label{SectionII}

\subsection{Typical strength of BWM signal}

In an astrophysical context, BWM typically occur in burst events
which are accompanied by considerable amount of mass or radiation
ejected in an asymmetric fashion. A simple formula for estimating
the characteristic amplitude of BWM  \citep{Braginsky1987} reads:
\bea \label{simpleformula} h^{\rm
mem}\sim\frac{r_g}{r}\left(\frac{v}{c}\right)^2,
 \ena
where $r_g$ and $v$ are the Schwarzschild radius and velocity of
aspherically  ejected parts respectively, and $r$ is the distance
to the source of the burst. The effect is maximal when the mass is
ejected at maximal speed, i.e \textit{c}, corresponding to
ejection of photons or gravitons. In the evaluations below we
shall work in units $[G=c=1]$.

An obvious source of GW BWM is a core-collapse  supernova. In a
typical scenario, the value of asymmetrically radiated energy is
$\Delta E_{\rm rad} \leq10^{-3}M_{\odot} $ (inferred from the
velocities of neutron stars, see e.g. \cite{Nazin1997}), leading
to an estimate $ h_{\rm 1Mpc}\leq 10^{-22}$, for the typical BWM
strain for a core-collapse supernova at a distance of $1~{\rm
Mpc}$ from the observer. In fact, these estimates are
overoptimistic (see \cite{Ott2009}). Taking into account the
expected event rate of core-collapse supernovae of a few per
century in a  Milky Way type galaxy, one can conclude that the
sensitivity of pulsar timing measurements will not allow the
detection of such a signal (see Section \ref{SectionIV} for a
discussion of the signal-to-noise ratio).

Another interesting class of burst  events are black hole (BH)
mergers \citep{Favata2009}. Depending on the angular momentum of
the merging BHs, up to several percent of the total mass of the
system can be radiated non-spherically during the burst
\citep{Reisswig2009}. The masses of BHs range from stellar to
several billion solar masses. However, from (\ref{simpleformula}),
we conclude that the most interesting candidates in the present
context are the mergers of super massive black hole (SMBH)
binaries, with typical masses $M_{\rm SMBH}\geq10^8M_{\odot}$.
Assuming that $10\%$ of the mass is radiated during the burst, for
the BWM strain at a distance of $1~{\rm Gpc}$ due to a SMBH merger
with total mass $10^8M_{\odot}$, we arrive at an estimate $h_{\rm
1Gpc} \simeq 10^{-15}$. As we shall show below, this strain is
within the reach of pulsar timing measurements.

This estimate can be made  more precise. The BWM amplitude from SMBH merger is given by Eq. (5) from
\cite{Favata2009}:
\begin{eqnarray}
h^{mem}=\frac{\eta M h}{384\pi
r}\sin^2\theta(17+\cos^2\theta),~~~{\rm where}~
h=\frac{16\pi}{\eta}\left(\frac{\Delta E_{\rm rad}}{M}\right).
\label{favataeq5}
\end{eqnarray}
In the above expression, $M=M_1+M_2$ is the total mass of SMBH,
$\eta=\frac{M_1+M_2}{M^2}$ and $\theta$ is the angle the binary
angular momentum and the line of sight. This formula can be
rewritten in the form:
\begin{eqnarray}
h^{mem}=\frac{\Delta E_{\rm rad}}{24
r}\sin^2\theta(17+\cos^2\theta).
\end{eqnarray}
Averaging over  $\theta$ yields:
\begin{eqnarray}
\label{amplitude} <h^{\rm mem}>=\frac{69}{8}\frac{\Delta E_{\rm
rad}}{24 r}\approx \frac{\Delta E_{\rm rad}}{3r}.
\end{eqnarray}
Estimates of $\Delta E_{\rm rad}$ can be obtained from
calculations of \cite{Reisswig2009} which give $\Delta E_{\rm
rad}$ from 3.6$\%$ to 10$\% ~M$. In the present work, we shall use
the mean value of $\Delta E_{\rm rad}=7\cdot10^{-2}~M$.  For
equal-mass SMBH system with $M_1=M_2=m=10^8~M_{\odot}$)  and  zero
orbital eccentricity at a distance of $r=1~\mathrm{Gpc}$, we
obtain:
\begin{eqnarray}
 h^{\rm mem}=
 5\cdot10^{-16}\left(\frac{m}{10^8~M_{\odot}}\right)\left(\frac{1~\mathrm{Gpc}}{r}\right).
 \label{BWMh}
\end{eqnarray}

At this point it is instructive  to compare the typical strain
associated with the BWM signal from SMBH merger with the strain
amplitude associated with the inspiral phase prior to the SMBH
merger. The characteristic strain amplitude associated with the
inspiral phase is given by \citep{Schutz2009} \bea
\label{inspiralh} h^{\rm insp}\approx
10^{-15}\left(\frac{m}{10^8~M_{\odot}}\right)\left(\frac{\rm
1~Gpc}{r}\right), \ena for the inspiral of SMBH of comparable
masses. Although the characteristic strains associated with the
BWM and the inspiral have comparable amplitudes, they lead to
different timing residual signatures. In a nutshell, the
qualitative difference in the residuals arises because the BWM
signal is characterized by a permanent offset of GW field which
leads to linear growth of the residuals with time, whereas the
inspiral signal has a quasi-periodic time varying GW field and consequently does
not lead to  linear growth of the residuals (see Section
\ref{SectionIII}).

In order to gain further insight, it  is helpful to present a
 heuristic argument for concentrating our effort on the signature of BWM
signal in pulsar timing residuals instead of the inspiral signal
although both have the comparable characteristic strains. We refer
the reader to Section \ref{SectionIV} for a rigorous calculation,
however for the purpose of order-of-magnitude estimation, it is
reasonable to assume that the signal-to-noise ratios (SNR)
associated with PTA measurement of BWM and inspiral signal are
related as:
$$
\frac{SNR_{\rm {BWM}}}{SNR_{\rm{insp}}} \sim \left(\frac{R_{\rm
{BWM}}}{R_{\rm {insp}}}\right),
$$
where $R_{\rm {BWM}}$, $R_{\rm {insp}}$ are timing residuals
associated with the BWM and inspiral signal, respectively. The BWM
residuals grow linearly with time $R_{\rm BWM}\sim h^{\rm
mem}T_{\rm obs}$ (see Section \ref{SectionIII}), where $T_{\rm
obs}$ is the total duration of observation. In the case of an
inspiral signal due to its quasi-periodic nature the residuals
oscillate with a characteristic amplitude $R_{\rm insp}\sim h^{\rm
insp}/\omega_{\rm insp}$, where $\omega_{\rm insp}\sim \left[
7.5\cdot 10^{3} \left( M/10^8M_{\odot} \right){\rm sec}
\right]^{-1}$ is the characteristic frequency associated with the
inspiral (which is the frequency at the Last Stable Circular
Orbit). Thus, assuming $T_{\rm obs} = 10~\rm{yrs}$, the expected
ratio is: \bea \frac{SNR_{\rm {BWM}}}{SNR_{\rm{insp}}}
\sim\frac{h^{\rm mem}T_{\rm obs} \omega_{\rm insp}}{h^{\rm
insp}}\sim T_{\rm obs} \omega_{\rm insp}\sim 2\cdot10^4. \nonumber
\ena For this reason, in what follows, we shall ignore the
possible contribution to the pulsar timing signal coming from
inspiral of the SMBH binary.

\subsection{Assessment of the event rate of SMBH mergers}

The crude estimation above shows that the amplitude of BWM from
SMBH mergers is within the expected reach of pulsar timing
measurements. Thus, the prospect of detecting the BWM signal from
SMBH binary mergers crucially depends on the expected event rate.
A typical pulsar timing measurement lasts for a period of $T_{\rm
obs} = 10~\rm{yrs}$. For this reason, if the event rate in a
volume of $1~\rm{Gpc}^3$ is larger or comparable to $0.1~{\rm
events~yr^{-1}}$, it is likely that a pulsar timing measurement
would be able to detect such an event.

The  event rate of SMBH mergers remains very uncertain. Various
theoretical models predict rates that differ by 2-3 orders of
magnitude \citep{LISA,volonteri2006}. The SMBH merger rate
primarily depends on two factors both of which need further study.
The first factor is the number of merging galaxies that contain
SMBH in the mass range of our interest. The second factor is the
fraction of galaxy mergers that leads to SMBH mergers. The galaxy
merger rate at redshifts $z<1$ is around one per year (see
\cite{LISA} and references therein). Therefore, simply assuming
that each of the merging galaxies contains an SMBH and that these
SMBH coalesce in each galaxy merger, the SMBH merger rate should
also be around one per year. Although very crude, these estimates
indicate that the SMBH merger rate is in the right ball-park to be
detected in pulsar timing measurements.

The event rate for SMBH  mergers in the mass range $10^7~M_{\odot}
< M < 10^9~M_{\odot}$ was calculated to be
$0.4~\rm{events}~\rm{yr^{-1}}$ in \citep{Sesana2004}. These
mergers occur for redshifts $z<4$, with maximum event rate at
$z\sim 2$. In addition, their calculations show that at least over
$20\%$ of SMBH mergers have mass ratios larger than $0.2$. A
concordant number for the expected event rate, of $0.1~\rm
events~yr^{-1}$ for SMBH merger at redshifts $z<1$ with total mass
$M\sim 10^{8}~M_{\odot}$ comes from \citep{Enoki2004}. Moreover,
as suggested by Fig.~6b in \citep{Enoki2004}, the event rate for
SMBH mergers with masses $10^7~M_{\odot}<M<10^8~M_{\odot}$ can be
about $1~\rm events~yr^{-1}$, albeit coming primarily from a
larger redshift $z\sim3$. Along with the theoretical estimates of
the SMBH merger event rates, recent studies \citep{Conselice2009}
suggest $R_g\sim ~10^{-3} ~\rm{events}~{Gpc}^{-3}yr^{-1}$ for the
rate of major galaxy mergers. Assuming that each major galaxy
merger is associated with SMBH binary coalescence, both numerical
estimations and observational data allows us to conclude that  a
rate of one event per ten years of observations within a redshift
of $z<0.5$ is not unrealistic.


\section{Signature of bursts with memory in pulsar timing residual}
\label{SectionIII}

\subsection{Pulsar timing residual due to a BWM}

A gravitational wave propagating between the pulsar and observer
leads to a modulation in the frequency of the observed pulsar
signal given by \citep{Estabrook1975,Sazhin1978,Detweiler1979}
\bea \frac{\Delta \nu}{\nu_0} = \frac{1}{2c}\int\limits_{0}^{D}
d\lambda \left.\left(e^ie^j \frac{\partial h_{ij}}{\partial
t}\right)\right|_{\rm path}, \label{Deltanu} \ena where $\nu_0$ is
the unperturbed pulsar frequency in the absence of gravitational
waves and $\Delta\nu(t) = \nu(t)-\nu_0$ is the variation of the
observed pulsar frequency due to the presence of a gravitational
wave characterized by the field $h_{ij}$. $D$ is the distance from
the pulsar to the observer and $c$ is the speed of light (below we set $c=1$). The
expression in the brackets is evaluated along the light ray path
from the pulsar to the observer, with the integration variable
$\lambda$ being the distance parameter along this path, and $e^i$
being the spatial unit vector along the path. The unperturbed
light ray path is given by \bea t(\lambda) = t -
{\lambda},~~~x^i(s) = x^i_O - e^i\lambda,
\label{trajectory} \ena where $t$ is the time of observation and
$x^i_{\rm O}$ is the position of the observer. Without loss of
generality we set $x^i_O=0$ by choosing a spatial coordinate
system with the observer at origin.

The pulsar timing measurements primarily measure the timing
residuals, i.e.~the difference between the actual pulse arrival
times and times predicted from a spin-down model for a pulsar
\citep{Detweiler1979}. The timing residual $s(t)$, accumulated
during a time interval of length $t$ beginning from an initial
time $t_{\rm in}=0$,  due to the presence of a gravitational wave
can be calculated from the expression for the frequency modulation
(\ref{Deltanu}) in the following way \bea s\left(t\right) =
\int\limits^{t}_{0} d\tau \frac{\Delta\nu(\tau)}{\nu_0}.
\label{Residual} \ena the residual $s\left(t\right)$ has
the dimensions of time and is customarily measured in nanoseconds.

In our analysis we shall assume that the gravitational wave source
is sufficiently far from the observer on Earth in comparison to
the distance $D$ between the observer and the pulsar, so as to
treat the GW in the plane wave approximation. In this case, the GW
incoming from the direction given by the unit vector $n_i$ can be
presented in the form \bea h_{ij} \left( x^i,t \right) =
h_{+}\left(t - n_ix^i\right) p_{ij}^{+} +
h_{\times}\left(t - n_ix^i\right)  p_{ij}^{\times} ,
\label{metric1} \ena where $h_+$ and $h_\times$ are the amplitudes
corresponding to  two linear polarization states of a GW. The linear polarization tensors
$p_{ij}^{+}$ and $p_{ij}^{\times}$
can be expressed in terms of two mutually orthogonal unit vectors $l_i$
and $m_i$ lying perpendicular to the direction of the wave propagation
$n_i$ as follows $p_{ij}^{+}=\left(l_il_j-m_im_j\right)$
and $p_{ij}^{\times}=\left(l_im_j+m_il_j\right)$. The
various projection terms that are encountered in evaluating
expressions (\ref{Deltanu}) and (\ref{Residual}) can be written as
\bea n_ie^i = \mu,~~~p_{ij}^{+}e^ie^j =
(1-\mu^2)\cos{2\phi}, ~~~p_{ij}^{\times}e^ie^j =
(1-\mu^2)\sin{2\phi}, \label{projections} \ena
in terms of angular variables $\mu = \cos{\theta}$ and $\phi$
(see, e.g., \cite{BPPP2008}). Here $\theta$ is the angle between
the direction of GW propagation $n^i$ and the direction from the
pulsar to the observer $e^i$. Angle $\phi$ is the azimuthal angle of
$e^i$ with respect to the principal direction $l_i$ and $m_i$
characterizing the GW polarization tensor, i.e. the angle between vector $l_i$ and the projection of vector $e^i$ onto $l_im_i$-plane.

The expression for frequency modulation (\ref{Deltanu}) can be
integrated for the  GW given in the form (\ref{metric1}) exactly.
Taking into account the projection factors (\ref{projections}), we
get \bea \frac{\Delta \nu(t)}{\nu_0} = && \frac{1}{2}~
(1+\mu)~ \left\{  \left[ h_{+}\left(t \right)\cos{2\phi} +
h_{\times}\left(t\right)\sin{2\phi}  \frac{}{}\right] -
\right. \nonumber \\
&& ~~~~~~~~~~~~~~\left.  - \left[
h_{+}\left(t -{D}\left(1-\mu\right)\right)\cos{2\phi} +
h_{\times}\left(t-D\left(1-\mu\right)\right)\sin{2\phi}
\frac{}{}\right]
 \right\}.
\label{Deltanu2} \ena The above expression has a clear physical
interpretation,  according to which the variation in frequency is
directly proportional to the difference between the GW field
strength at the place and time of observation (terms in the first
square bracket) and its strength at the place and time of signal
emission (terms in the second square brackets). In specific case
of a BWM signal that reached the Earth during time span of pulsar
observation, the second term will naturally vanish since the strength of the
signal at the place and time of emission would have been equal to zero.
On the other hand, it can be seen from (\ref{Residual}) and (\ref{Deltanu2})
that if BWM front crossed the pulsar neighborhood at or before the moment of
emission the second term would give rise to linear rising term in
prefit timing residuals. This linear term would be absorbed after fitting
procedure (see below) and so the second term can be
neglected in this case also. For these reasons, below we shall omit the second term in
(\ref{Deltanu2}).

Using expression (\ref{Deltanu2}) and neglecting the terms
proportional to the GW field strength at emission, we arrive at
the expression for the timing residual \bea s(t) = \frac{1}{2}~
(1+\mu)~\left\{ \left( \int\limits_0^{t}d\tau~h_{+}\left(\tau
\right)\right)\cos{2\phi} + \left(
\int\limits_0^{t}d\tau~h_{\times}\left(\tau
\right)\right)\sin{2\phi}
 \right\}.
\label{Residual2}
\ena

We shall model the  BWM in a simple analytical manner (that would
be applicable in a wide range of situations) as a  step-function
signal \bea
 h_{+}(t) =h^{mem} \Theta(t-t_B),~~h_{\times}(t) = 0,
\label{Theta} \ena where $t_B$ is the time the BWM signal reaches
the observer on Earth. The function $\Theta(x)$ is the Heaviside
step function \bea \Theta(x) = \left\{\begin{array}{l} 0,~~x\leq0
\\ 1,~~x>0 \end{array} \right. \label{Theta1} \ena

From (\ref{Residual2}) and (\ref{Theta}) we find the  prefit
timing residuals:
\bea s\left(t\right)_{\rm prefit} =
\frac{1}{2}h^{\rm mem} (1+\mu)\cos{2\phi} ~\left(t-t_{\rm B}\right) \Theta\left(t-t_{\rm B}\right),
\label{Residual3} \ena
 Postfit residuals are obtained by
the removal of linear and quadratic trends from prefit ones to
take into account a priori unknown values of the pulsar period and
its first derivative. \bea s\left(t\right) =
\frac{1}{2}h^{\rm mem} (1+\mu)\cos{2\phi}
~\mathcal{I}(t), \label{Residual3} \ena where
$\mathcal{I}\left(t\right)$ is given by the expression
\bea \mathcal{I}\left(t\right) = \left(t-t_{\rm B}\right)
\Theta\left(t-t_{\rm B}\right) - \mathcal{I}_{\rm quad}(t).
\label{IIntegral} \ena

\begin{figure}
\begin{center}
\includegraphics[width=8cm]{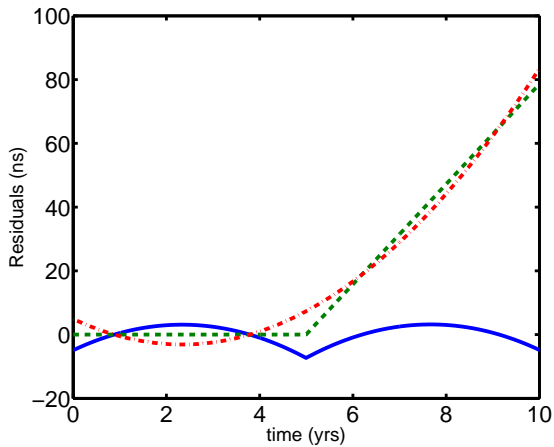}
\end{center}
\caption{The solid line shows the expected timing residual  due to
a BWM with $h^{\rm mem}=5\cdot10^{-16}$ and $t_B=5~\rm{yrs}$, in a
pulsar with directional angles $\mu=\phi = 0$, during an
observation spanning $10~\rm{yrs}$, after subtraction of the
quadratic fitting term. The dashed and the dashed-dotted lines
show the BWM residual signal before subtraction and quadratic
subtraction terms, respectively.}\label{figure1}
\end{figure}

\noindent The quadratic subtraction term $\mathcal{I}_{\rm
quad}(t)$ can be written in a general form \bea \mathcal{I}_{\rm
quad}\left(t\right) = a \left(t-t_{\rm B}\right)^2 + b
\left(t-t_{\rm B}\right) + c. \label{Iquad1} \ena The
coefficients $a$, $b$ and $c$ can be evaluated by minimizing the
integral $\int_{0}^{T_{\rm obs}} dt ~\mathcal{I}^2\left(t\right)$ (where $T_{\rm obs}$ is the total duration of observation)
with respect to these variables. For example, in the case of a BWM
signal occurring at $t_B = T_{\rm obs}/2$, these coefficients take a
particularly simple form \bea a = \frac{15}{16T_{\rm obs}}, ~~~ b =
\frac{1}{2}, ~~~ c = \frac{3T_{\rm obs}}{64}. \label{Iquad2} \ena The
signature in the pulsar timing residuals of such model BWM is
shown in  Fig.~\ref{figure1}.

In the following subsections, we shall assess the sensitivity of PTA observations to the BWM signal, and estimate the expected rate of observable events.

\subsection{Signal-to-noise ratio}
\label{SectionIV}

In order to quantify the detection ability of PTA, in the present
section we  introduce and evaluate the signal-to-noise ratio for a
BWM event. The observed residuals in a pulsar timing array can be
presented in the form\bea R_{\alpha}\left(t_i\right) =
s_{\alpha}\left(t_i\right)+n_{\alpha}\left(t_i\right), \ena where
the index $\alpha=1,..,N_\alpha$ marks the residuals measured for
the $\alpha^{\rm th}$ pulsar, and the index $i=1,..,N_t$ is the
number of the measurement. $N_{\alpha}$ and $N_t$ are the number
of pulsars in the timing array and the number of time observations
per individual pulsar, respectively. $s_{\alpha}\left(t_i\right)$
is the part of  residuals due to BWM (Eq. (\ref{Residual3})), and
$n_{\alpha}\left(t_i\right)$ is the noise.

We shall assume that  $n_{\alpha}\left(t_i\right)$ is gaussian
stationary white noise, uncorrelated for different pulsars. Under
these assumptions, the noise correlation function will have the
form \bea
\overline{n_{\alpha}\left(t_i\right)n_{\beta}\left(t_j\right)} =
\sigma_n^2({\alpha})\delta_{ij}\delta_{\alpha\beta},
\label{noisespectrum} \ena  where $\delta_{\alpha\beta}$ and
$\delta_{ij}$ are the Kronecker deltas. The residuals
$s_{\alpha}(t)$ can be conveniently rewritten as \bea
s_{\alpha}\left(t_i\right) = h^{\rm
mem}f\left(\mu_{\alpha},\phi_{\alpha}\right)\mathcal{I}\left(t_i\right),
\label{signalfactorization} \ena where
$\mathcal{I}\left(t_i\right)$ is the common part of the signal for
all pulsars (\ref{IIntegral}), and
$f\left(\mu_{\alpha},\phi_{\alpha}\right)$ is the part of the
residuals  that depends on the orientation angles of the pulsars
$(\mu_\alpha,\phi_{\alpha})$ \bea
f\left(\mu_{\alpha},\phi_{\alpha}\right) =  \frac{1}{2}
(1+\mu_{\alpha})\cos{2\phi_{\alpha}}. \ena

The common way to extract a signal with a known form from a
gaussian stationary noise is by using a matched filter
\citep{Schutz2009}. The signal-to-noise ratio $\rho$ attainable
using the matched filter can be written in terms of the power of
the noise $ \sigma_n^2({\alpha})$ and the expected signal
$s_{\alpha}\left(t_i\right)$ \bea \rho^2 =
\sum_{\alpha=1}^{N_\alpha} \left(
\frac{1}{\sigma_{n}^2(\alpha)}\sum_{i=1}^{N_t} s^2_{\alpha}
\left(t_i\right) \right). \label{SNR1} \ena Note that the above
expression for signal-to-noise ratio is derived in time-domain
variables which is more convenient in our case.  Using
(\ref{signalfactorization}), $\rho$ can be rewritten in a
factorized form \bea \rho^2 = \left({h^{\rm mem}}\right)^2
N_tN_\alpha \left( \frac{1}{N_{\alpha}} \sum_{\alpha=1}^{N_\alpha}
\frac{f^2\left(\mu_{\alpha},\phi_{\alpha}\right)}{\sigma_{n}^2(\alpha)}
\right) \left( \frac{1}{N_t} \sum_{i=1}^{N_t} \mathcal{I}^2
\left(t_i\right) \right). \label{SNR2} \ena  Assuming a near
isotropic sky coverage for the pulsar timing array and a similar
noise level for all the pulsars $\sigma_n$, the terms in the first
bracket can be calculated  as a sky-average value \bea
\frac{1}{N_{\alpha}} \sum_{\alpha=1}^{N_\alpha}
\frac{f^2\left(\mu_{\alpha},\phi_{\alpha}\right)}{\sigma_{n}^2(\alpha)}
\approx \frac{1}{4\pi\sigma_n^2}\int d\mu d\phi~ f^2(\mu,\phi) =
\frac{1}{6\sigma_n^2}. \label{approximation1} \ena The terms in
the second bracket in expression (\ref{SNR2}) can be approximated
by an integral \bea \frac{1}{N_t} \sum_{i=1}^{N_t} \mathcal{I}^2
\left(t_i\right) \approx \frac{1}{T_{\rm
obs}}\int\limits_0^{T_{\rm obs}} dt~\mathcal{I}^2(t).
\label{approximation2} \ena This integral can be calculated
explicitly if the form of $\mathcal{I}_{\rm quad}$ in Eq.
(\ref{IIntegral}) is known. For example, in the case of
$t_B=T_{\rm obs}/2$, the form of $\mathcal{I}_{\rm quad}$ is given
by expressions (\ref{Iquad1}) and (\ref{Iquad2}) and  \bea
\int\limits_0^{T_{\rm obs}} dt~\mathcal{I}^2(t) = \frac{T_{\rm
obs}^3}{3072}, \label{IntegralValue} \ena It is convenient to
introduce  a dimensionless quantity $\iota$ \bea \iota =
\sqrt{\frac{\int\limits_0^{T_{\rm obs}} dt
~\mathcal{I}^2(t)}{T_{\rm obs}^3/3072}}, \ena  which is about one
for
 $0<t_{\rm B}< T_{\rm obs}$.

In the future PTA observations $N_t=250$, $N_\alpha=20$, $T_{\rm
obs}=10~{\rm yrs}$, $\sigma_n = 100 ~{\rm ns}$ (that is
the level of sensitivity currently achieved for a number of
pulsars \citep{Verbiest2009} ). Taking these numbers along with a
signal strength $h^{\rm mem} = 10^{-15}$ as guidelines and using
approximations (\ref{approximation1}) and (\ref{approximation2}),
the expression for signal-to-noise ratio can be rewritten in the
form \bea \rho = 1.64 \left[ \iota \left(\frac{h^{\rm
mem}}{10^{-15}}\right)
\left(\frac{N_t}{250}\right)^{\frac{1}{2}}\left(\frac{N_\alpha}{20}\right)^{\frac{1}{2}}
\left(\frac{T_{\rm obs}}{10{\rm yrs}}\right) \left(\frac{100{\rm
ns}}{\sigma_n}\right) \right]. \label{snr} \ena


\subsection{Expected event rate}
\label{SectionIV} Equation  (\ref{snr}) shows the practical
possibility of using pulsar timing measurements to detect
individual GW bursts that accompany SMBH mergers. As can be seen
from Eq. (\ref{snr}), the BWM with an amplitude of $h^{\rm
mem}\sim (1.5-2)\cdot 10^{-15}$ can be detected in a 10-years run
of PTA observations with a signal-to-noise ratio of $\sim 3$. Eq.
(\ref{BWMh}) indicates that such a signal would be produced by the
coalescence of two SMBH with equal masses around $3.5\cdot 10^8~
M_{\odot}$ at a distance of 1 Gpc.

It is instructive to substantiate this rate  by properly taking
into account the contribution to event rate from SMBH mergers at
various cosmological distances (redshifts). The rate of GW bursts
of sufficient strength from coalescing SMBH coming from redshifts
$z<z_{\rm lim}$, $\dot N(z_{\rm lim})$, ($[\dot N(z_{\rm
lim})]=\rm{yr}^{-1}$) can be calculated from  the number density
of SMBH mergers $n(z)$ with masses $M_{\rm BH}>M_{\rm lim}(z)$,
($[n (z)] = ~\rm{Mpc}^{-3}$) and the characteristic merging
frequency for a SMBH, $\eta(z)$, $[\eta]=~\rm{yr}^{-1}$. The event
rate $\dot{N}(z_{\rm lim})$ of bursts with sufficient strength is
\bea
 \dot N(z_{\rm lim})=\int_0^{z_{\rm lim}}n_0(1+z)^3\frac{\eta(z)}{1+z}4\pi
 r^2dr\,,
\label{rate1}
\ena
 where $r$ is the metric distance determined from
 \bea
\frac{dr}{dz}=\frac{c}{H_0}\frac{1}{\sqrt{\Omega_m(1+z)^3+\Omega_{\Lambda}}}
 \label{drdz}
 \ena
in a flat $\Lambda$CDM cosmological model. Here $H_0$ is the
present-day value of the Hubble parameter and $\Omega_\Lambda$,
$\Omega_m$ stand for the cosmological constant and matter energy
content in units of the critical density,
$\Omega_m+\Omega_\Lambda=1$. For numerical estimations below we
adopt the standard values $H_0=72~\rm km~ s^{-1} Mpc^{-1}$,
$\Omega_m=0.3$ and $\Omega_\Lambda=0.7$.

The local comoving density of SMBH $n_0$  can be estimated using the
integral SMBH mass-function from the study of \cite{Caramete2009}:
\bea
\phi(M_{\rm BH})\equiv N(M>M_{\rm BH})=7\cdot10^{-2}\left(\frac{M_{\rm
BH}}{10^7~M_{\odot}}\right)^{-2}~\rm Mpc^{-3}\,.
\label{mass-function}
 \ena

The local SMBH merging rate $\eta_0$ can be estimated, assuming
that each SMBH with the mass greater than $10^8$~M$_\odot$ has
undergone at least one merging in the past, from the local density
of SMBH $n_0$ and the specific merging rate per unit volume
$\mathcal{R}_0$. Using the common assumption (e.g.
\cite{Enoki2004}) that each major galaxy merging is associated
with a SMBH coalescence,  the specific SMBH merging rate can be
obtained from the analysis of major mergings of galaxies with
stellar mass $M_*>10^{10}$~M$_{\odot}$ (Conselice et al. 2009),
$\mathcal{R}_0=10^6~\rm Gpc^{-3}Gyr^{-1}$. \bea
\eta_0=\frac{\mathcal{R}_0}{\rho_0}\approx\frac{10^{-12}~\rm
Mpc^{-3}yr^{-1}}{10^{-2}~\rm Mpc^{-3}}=10^{-10}~\rm yr^{-1}
\label{eta}\,. \ena In this estimation we have used the local
density of massive galaxies $\rho_0\sim 0.01$~Mpc$^{-3}$. The
obtained value agrees with the frequency of major galaxy mergers
from (Conselice et al. 2009) $\eta_0=(1-2)\cdot 10^{-10}~ {\rm
yr}^{-1}$.

We are interested only in the SMBH merger events that could be registered with PTA. For this reason, more distant SMBH mergings should be more massive to be detected, and the corresponding minimal detectable mass scales as
$$
M_{\rm lim}(z)\propto r(1+z)\,.
$$
Setting the reference point at $r_0=1$~Gpc ($z\sim 0.2$) we get the corresponding reference values for SMBH mass $M_{0}$ and  local comoving density of SMBH $n_0$:
\bea
M_{0}=3.5\cdot10^8~M_{\odot},
~~ n_0=6\cdot10^{-5}~\rm{Mpc}^{-3}.
\label{mono}
\ena
In terms of the reference values, the minimum detectable mass can now be expressed in the form:
\bea
M_{\rm lim}(z)=M_0\left(\frac{r(1+z)}{r_0(1+z_0)}\right).
\label{refmass}
\ena
Using Eqs. (\ref{mass-function}), (\ref{mono}) and (\ref{refmass}), we arrive at
\bea
n(z)=n_0\left(\frac{M_0}{M_{\rm lim}(z)}\right)^2 =
6\cdot10^{-5}\left(\frac{r_0(1+z_0)}{r(1+z)}\right)^2
\approx10^{-4}\left(\frac{r_0}{r(1+z)}\right)^2\,.
\label{density_z}
\ena

Substituting Eq. (\ref{density_z}) into Eq. (\ref{rate1}) we find
\bea
\dot N(z_{\rm lim}) =
4\pi\cdot10^{-14}r_0^2\int_0^{z_{\rm
lim}}dr=4\pi\cdot10^{-14}r_0^2\int_0^{z_{\rm lim}}\frac{dr}{dz}dz
\label{rate2}
\ena
After substituting (\ref{drdz}) into above equation, we finally obtain
\bea
\dot N(z_{\rm lim})&=&4\pi\cdot10^{-14}\frac{r_0^2c}{H_0}\int_0^{z_{\rm
lim}}\frac{1}{\sqrt{\Omega_m(1+z)^3+\Omega_{\Lambda}}}dz
\nonumber \\ &=& 5\cdot10^{-4} \int_0^{z_{\rm
lim}}\frac{1}{\sqrt{\Omega_m(1+z)^3+\Omega_{\Lambda}}}dz.
\label{rate3}
\ena
We set our limiting redshift to $z_{\rm lim}=5$, bearing in mind that
SMBH assembly increasingly occurs around $z\sim 3$
(see, e.g., \cite{Enoki2004}). In that case the value of the integral
in Eq. (\ref{rate3}) is close to 2 and the detection rate is
\bea
\dot N(z_{\rm lim}=5)\approx10^{-3}~\rm yr^{-1}\,.
\label{rate4}
\ena

Taken at face value, this rate seems to be fairly low, but we can
increase it significantly if we take into account the strong
redshift dependence of the specific major galaxy merging rate
$\eta(z)=\eta_0(1+z)^\beta$. With this factor the integral in Eq.
(\ref{rate3}) takes the approximate value: \bea
\int_0^5\frac{1}{\sqrt{\Omega_m(1+z)^3+\Omega_{\Lambda}}}(1+z)^{\beta}dz
\approx1.25\cdot10^{0.58\beta} \label{eta_approx}\,. \ena This
approximation slightly underestimates the integral for $\beta\sim
0$. The analysis of (Conselice et al. 2009) suggests $\beta=2-3$.
Adopting $\beta = 2$ we obtain the SMBH detection rate from the
limiting redshift $z_{\rm lim}=5$ \bea \dot N(z_{\rm
lim}=5)\approx 6\cdot10^{-4+0.58\beta}~\rm yr^{-1}\approx
10^{-2}~yr^{-1}\,. \label{rate5} \ena

Our figure of merit is the number of detections with $\rm SNR$
above a fixed threshold value $\rho$ in a PTA observation with
given sensitivity (i.e. the rms of timing residuals $\sigma_n$)
over the total observing time $T_{\rm obs}$. This number is
determined by the product $\dot N\times T_{\rm obs}$.
The detection rate $\dot N\left(z_{\rm lim}\right)\propto
n_0(M_0)\propto M_0^{-2}\propto \left(h^{\rm mem}\right)^{-2}$,
where $h^{\rm mem}$ is the signal amplitude that triggers the
detector at a given SNR level $\rho$ (see Eq. (\ref{snr})), i.e.
$h^{\rm mem}\propto \rho N_t^{-1/2}N_\alpha^{-1/2}T_{\rm
obs}^{-1}\sigma_n$. So finally, we find the number of detections
in observations with duration $T_{\rm obs}$ is \bea N\simeq
10^{-1}\left(\frac{N_t}{250}\right)\left(\frac{N_\alpha}{20}\right)
\left(\frac{T_{\rm obs}}{10~{\rm yrs}}\right)^3
\left(\frac{100~{\rm
ns}}{\sigma_n}\right)^2\left(\frac{3}{\rho}\right)^2\,.
\label{number_of_detection} \ena It can be seen that the expected
number of detections is very sensitive to the duration of
observations and the rms noise level.


\section{Conclusions}
\label{SectionV}

We have shown that future pulsar timing measurements will be
capable of detecting individual gravitational wave bursts that
should accompany SMBH mergers. A GW burst with memory with an
amplitude of $\sim (1.5-2)\cdot 10^{-15}$ leaves the
characteristic imprint in the pulsar timing residuals and can be
detected in a 10-years run of Pulsar Timing Array observations
with the present-day characteristics at a signal-to-noise ratio of
$\sim 3$. Such a signal is expected to be produced by the
coalescence of two SMBH with equal masses around $3.5\cdot 10^8
~M_{\odot}$ at a distance of 1 Gpc. We estimate the rate of SMBH
coalescences producing the BWM with such an amplitude to be around
a few hundredth per year. The number of detections of such GW
bursts from SMBH mergings by a PTA array with given
characteristics is given by Eq. (\ref{number_of_detection}) that
shows a strong dependence on the total time of observations and
the noise level in pulsar timing residuals. It is expected that
increase in the PTA sensitivity in the near future will allow
detection of BWM from coalescing SMBHs at a signal-to-noise ratio
more than three. Future radiotelescopes, especially SKA, will be
able to increase the sensitivity several times due to the decrease
in timing residual noise and the extension of number of pulsars in
pulsar timing program, thus making this method complementary to
the LISA space mission for detection of coalescing SMBHs with
masses $>10^{8}~M_{\odot}$ to which the sensitivity of LISA is
reduced.

Finally, it is worth pointing out that the analysis conducted  in
this paper is applicable for any source of BWM, not specifically
restricted to GW signals from SMBH mergers. As was mentioned
above, BWM are a generic feature of GW sources that emit energy in
an asymmetric fashion. These BWM events will leave an imprint in
pulsar timing residuals as long as they are sufficiently bright
$h^{\rm mem}\gtrsim 10^{-15}$.

When this paper was already submitted, two studies
  dealing with the influence of BWM on PTA appeared \citep{Seto2009,vHaasteren2009}.
 Results of \citep{vHaasteren2009} are based on  Bayesian analysis method and  give the same conclusions,
 results of  \citep{Seto2009} with different treatment of background noise  is somewhere more pessimistic.



\section*{Acknowledgements}
The authors thank Joseph Romano, Patrick Sutton, A.V. Zasov and
O.K. Sil'chenko for useful discussions. Also the authors thank Kip
Thorne for fruitful suggestions. The work is supported by RFBR
grants 07-02-00961, 07-02-01034 and 09-02-00922. This research has
made use of NASA's Astrophysics Data System.



\end{document}